\documentclass[5p]{elsarticle}

\usepackage{url} 
\usepackage{lineno,hyperref}
\usepackage{color}
\usepackage{graphicx}

\usepackage{aas_macros}
\usepackage{amsmath}
\usepackage{enumitem}

\journal{Astronomy and Computing}

\begin{document}

\title{Data Processing Software for Chandrayaan-2 Solar X-ray Monitor}

\author[1]{N. P. S. Mithun}
\ead{mithun@prl.res.in}
\author[1]{Santosh V. Vadawale}
\author[1]{Arpit R. Patel}
\author[1]{M. Shanmugam}
\author[1]{D. Chakrabarty}
\author[1]{Partha Konar}
\author[1]{Tejas N. Sarvaiya}
\author[1]{Girish D. Padia}
\author[1]{Aveek Sarkar}
\author[1]{Prashant Kumar}
\author[1]{Prashant Jangid}
\author[1]{Aaditya Sarda}
\author[1]{Manan S. Shah}
\author[1]{Anil Bhardwaj}

\address[1]{Physical Research Laboratory, Ahmedabad, India}

\begin{abstract}

\textit{Solar X-ray Monitor} (XSM) instrument of India's Chandrayaan-2 lunar
mission carries out broadband spectroscopy of the Sun in soft
X-rays. XSM, with its unique features such as low background, high time 
cadence, and high spectral resolution, provides the opportunity to characterize
transient and quiescent X-ray emission from the Sun even during low activity 
periods. 
It records the X-ray spectrum at one-second cadence, and the data
recorded on-board are downloaded at regular intervals along with that of
other payloads. During ground pre-processing, the XSM data is segregated,
and the level-0 data is made available for higher levels of processing at
the Payload Operations Center (POC). XSM Data Analysis Software (XSMDAS)
is developed to carry out the processing of the level-0 data to higher levels and
to generate calibrated light curves and spectra for user-defined binning
parameters such that it is suitable for further scientific analysis. A
front-end for the XSMDAS named XSM Quick Look Display (XSMQLD) is also
developed to facilitate a first look at the data without applying
calibration. XSM Data Management-Monitoring System (XSMDMS) is designed to carry out
automated data processing at the POC and to maintain an SQLite database with
relevant information on the data sets and an internal web application for
monitoring data quality and instrument health. All XSM raw and calibrated
data products are in FITS format, organized into day-wise files, and the
data archive follows Planetary Data System-4 (PDS4) standards. The XSM
data will be made available after a lock-in period along with the XSM Data
Analysis Software from ISRO
Science Data Archive (ISDA) at Indian Space Science Data Center(ISSDC). 
Here we discuss the design and implementation
of all components of the software for the XSM data processing and the contents of
the XSM data archive.
\end{abstract}


\maketitle

\section{Introduction}

The \textit{Solar X-ray Monitor} (XSM)~\citep{shanmugam20,mithun20} on-board the orbiter of the Chandrayaan-2 mission
performs spectroscopic observations of the Sun from a lunar orbit.
It provides measurements of the soft X-ray solar spectrum in the energy range of 1--15 keV.
These measurements are used to infer the temperature distribution, 
abundances, and evidence of non-thermal processes in the solar corona. Also, the XSM spectra along with the X-ray fluorescence 
spectra from the Moon obtained by \textit{Chandrayaan-2 Large Area Soft X-ray Spectrometer} 
(CLASS) \citep{radhakrishna20} are used to estimate elemental abundances on the lunar surface.

XSM offers spectral monitoring of the Sun in a broad energy band in soft X-rays with 
an energy resolution of better than 180 eV at 5.9 keV with a stable spectral performance 
over the wide range of X-ray flux from the Sun. It is designed to carry out these measurements with the 
highest time cadence so far of one second, over a broad range of solar activity, i.e. X-ray emission 
below A-class level up to X-class.
For this purpose, it employs a Silicon Drift Detector as the sensing device and includes a
filter wheel mechanism to automatically bring a beryllium filter in front of the detector when
the flux exceeds a certain threshold. The filter wheel also includes a radioactive
source (Fe-55) for in-flight calibration. The filter wheel is brought to the source position 
for calibration observations, whereas during regular observations of the Sun or background,
it is kept in the open position with a possibility of automatic change over to
the Be filter during intense solar flares. As the instrument is fix-mounted on the spacecraft, the angle
subtended by the Sun with the XSM axis varies during observations. 
There will also be occasions when the Sun moves out of the field of view (FOV) of the XSM or is 
occulted by the Moon, and the data acquired during such periods provide the measurement of background spectrum.

The Chandrayaan-2 mission was launched on 22 July 2019, and the orbiter reached the final lunar orbit in early September. The XSM began its observations on 12 September 2019 and has been acquiring data since then. The data recorded on-board is downloaded to the ground stations at regular intervals and the pre-processing 
is carried out at Indian Space Science Data Center (ISSDC), Bangalore. 
The data from the instrument has to then undergo several stages of processing 
such as incorporating the necessary calibration corrections to generate products for scientific use.  
For this purpose, a suite of software tools are developed 
and are made operational at the XSM Payload Operations Center (POC) located at 
Physical Research Laboratory (PRL), Ahmedabad. These software tools include a user-level data analysis software, a quick look software, and a host of
tools for the management of automated processing at the POC. In this article, we present the design and implementation of these software modules for the XSM and a description of
the data products and archive.

This article is structured as follows: Section~\ref{sec:procreq} gives an overview
of the XSM data and the processing requirements. Sections~\ref{sec:xsmdas} and
~\ref{sec:xsmqld} describe the XSM Data Analysis Software and XSM Quick Look Display,
respectively. The architecture of the automated processing chain at the POC is given in section~\ref{sec:xsmpocms}, and section~\ref{sec:dataproduct} provides
details of the data products and archive followed by a summary.

\section{XSM data and processing requirements}
\label{sec:procreq}

\subsection{On-board data management}

XSM detects individual X-ray photons that interact with its detector and measures
their energy in Analog to Digital Converter (ADC) value that corresponds
to the charge generated by the photon~(see \cite{shanmugam20} for more details). Hence, the basic data collected by the
onboard FPGA (Field Programmable Gate Array) based processing electronics is the ADC or PHA (Pulse Height Analysis) channel
of the individual photons or events. It is common to record this list of events
with the associated time stamps and PHA values and transfer it
to the ground where further analysis can be carried out. However, in the case of
XSM, given the wide range of incident X-ray rates expected from the Sun over
its various classes of activity, recording the complete event list would result
in a significant volume of data. Moreover, as the incident rate varies over several
orders of magnitude, the data rate will not remain constant, making it even more
challenging to handle. Thus, instead of storing the event-mode data, the XSM onboard
software generates a histogram of the 10-bit PHA values of the photons 
detected in one second duration and records this 1024 channel spectrum in a memory bank.

The XSM processing electronics then generates a data packet of 2048 bytes 
(format given in table~\ref{xsm_data_format}) every second
with this spectrum along with other housekeeping information and XSM clock time.
As shown in table~\ref{xsm_data_format}, counts in PHA channels that correspond to 
energies above $\sim$15 keV (PHA$>$961) are recorded together 
so that the remaining space in the data packet can include the instrument
housekeeping parameters such as the filter wheel position, various voltage levels,
detector temperature, and other instrument settings. Apart from the full spectrum, the onboard software registers the number
of events in three pre-defined PHA ranges at every 100 ms interval and 
these light curves are also included in the data packet. At the end of one second, the processing
electronics transfer the data packet from its one memory bank to the spacecraft data
handling (DH) system while starting to record the spectrum in the second memory bank.
The data handling system records the XSM data packet with additional header 
information, including the DH clock time in its solid-state recorder (SSR) along with the data from
all other instruments. The SSR data is played back to ground stations at
regular intervals considering the availability of onboard storage 
and ground station visibility.

 \begin{table}
 \caption{Format of the XSM data packet of size 2048 bytes. Counts in 
each channel of the spectrum have a width of two bytes.}
 \label{xsm_data_format}
 \begin{tabular}{l l}
 \hline
 \hline
 Byte range & Content \\
 \hline
 0-3    & Start bytes \\
 4-9    & XSM clock time \\
 10-13  & Packet sequence counter \\
 14-17  & Event trigger counter \\
 18-21  & Detected event counts \\
 22-55  & Housekeeping parameters\\
 56-115 & 100 ms light curves in 3 PHA ranges\\
 116-121& Header end bytes\\
 122-2043&  Spectrum for channel range 0-961  \\
 2044-2045& Total counts in channels 962-1022 \\
 2046-2047& Counts in channel 1023 (ULD) \\
 \hline
 \hline
 \end{tabular}
 \end{table}

\subsection{Ground pre-processing}

The raw frames downloaded from the spacecraft at the ground stations contain
data from all instruments that were operated during that session. The data from different
instruments are segregated during the first stage of ground pre-processing. In the next step of level-0
processing at ISSDC, each of the segregated XSM data packets with the DH header is assigned
the corresponding UTC stamp based on the DH clock time. Correlation between DH clock
time and UTC is derived from real-time telemetry samples correcting for the transmission
delay and other factors. 

Analysis and interpretation of the instrument science data
require other auxiliary information like the orbit and attitude of the spacecraft during the
observation. This auxiliary information derived from the spacecraft telemetry data and orbit
determination are provided in the form of SPICE kernels of orbit, attitude, and clock information
for the duration of the instrument data. The instrument data packets with the UTC stamps
in a binary (.pld) file along with this auxiliary information as SPICE kernels
constitute the level-0 data product for each download session.
It may be noted that the successive downloads, in general, will have an overlap of data
in order to ensure continuity.
The level-0 data sets are generated at ISSDC soon after the data download and are then accessible
from the POC for the higher levels of processing.

\subsection{Data processing requirements}

The objective of any data processing software is to generate calibrated science data
products from the raw instrument data. In the case of XSM,
these science data products are (i) spectrum that provides the count rates as a
function of energy over a specified time interval; and (ii) light curve which is
count rate as a function of time over a given energy range. These products
should be in formats compatible with standard X-ray spectral and timing analysis
software tools. The data acquired by XSM may include duration which are to be
removed for analysis such as the periods when the Sun is out of FOV of XSM
or when the detector temperature is not within favorable limits.
The light curves and spectra are to be generated for `Good Time Intervals' (GTI)
that avoid such periods of time. Also, the data analysis routines should incorporate the required
corrections, such as the instrument's gain to convert from raw PHA channels
to energy and for the change in effective area as a function of the incidence angle, 
which are determined from the ground calibration of the instrument~\citep{mithun20}.
Although the calibrated light curves and spectra thus generated for a standard energy range or time bins
will be useful, the users often require to generate them with specific
set of input parameters depending on their objectives. 
Hence, the processing software for generating the science products should be 
designed for distribution along with the raw data.

Raw FITS data files of the XSM should include the 
instrument data packets and other necessary auxiliary information 
such as the housekeeping and observation geometry parameters. 
Although the level-0 data from successive downloads usually have overlaps and
span arbitrary time durations, the FITS files generated from level-0 at the POC are
to be organized as day-wise files along with the corresponding standard calibrated
products. Further, as the data from the Chandrayaan-2 mission is planned to
be archived following Planetary Data System-4 (PDS4)\footnote{\url{https://pds.jpl.nasa.gov/datastandards/about/}} standards,
the data files are to be accompanied by respective  XML labels with the metadata,
which also should be created by the data processing software.

During the commissioning phase of the instrument, it is often useful to have
visualization of the data without any corrections as and when it is downloaded 
from the spacecraft.
For this purpose, a quick look software with a graphical user interface
is required. However, once the operations are regularized, it is more useful
to have a web application at the POC with all quick look plots including that of the calibrated
products. It would also be required to automate the processing at the POC and to
keep records of the data sets and their associated information. 
These activities at the POC also require dedicated software tools.

In order to cater to all these requirements, we have developed the following software
suites: (i) XSM Data Analysis Software (XSMDAS), (ii) XSM Quick Look Display (XSMQLD), and
(iii) XSM Data Management-Monitoring System (XSMDMS), details
of which are given in the subsequent sections.

\section{XSMDAS: XSM Data Analysis Software}
\label{sec:xsmdas}

The XSM Data Analysis Software (XSMDAS) is designed to process the level-0 data
to level-1 and level-2 data products. It is composed of individual modules that
cater to specific functionality and is meant to be used by the 
scientific users for the analysis of XSM data. All calibration information of the XSM required for the analysis are included in a calibration
database (CALDB), which is used by the XSMDAS modules. 
Details of the XSM data level 
definition, XSMDAS architecture and algorithms, and the XSM
CALDB are discussed in this section.

\subsection{Data levels and XSMDAS architecture}

XSM data are organized into two levels: level-1 that includes raw frames and computed auxiliary information 
organized as day-wise FITS files and level-2 that includes calibrated science products.  
The XSMDAS modules generate level-1 files from level-0 data and level-2 products
from the level-1 data. Details of the contents of the data files 
are discussed in section~\ref{sec:dataproduct}. 

\begin{figure}
\centerline{\includegraphics[width=0.95\columnwidth]{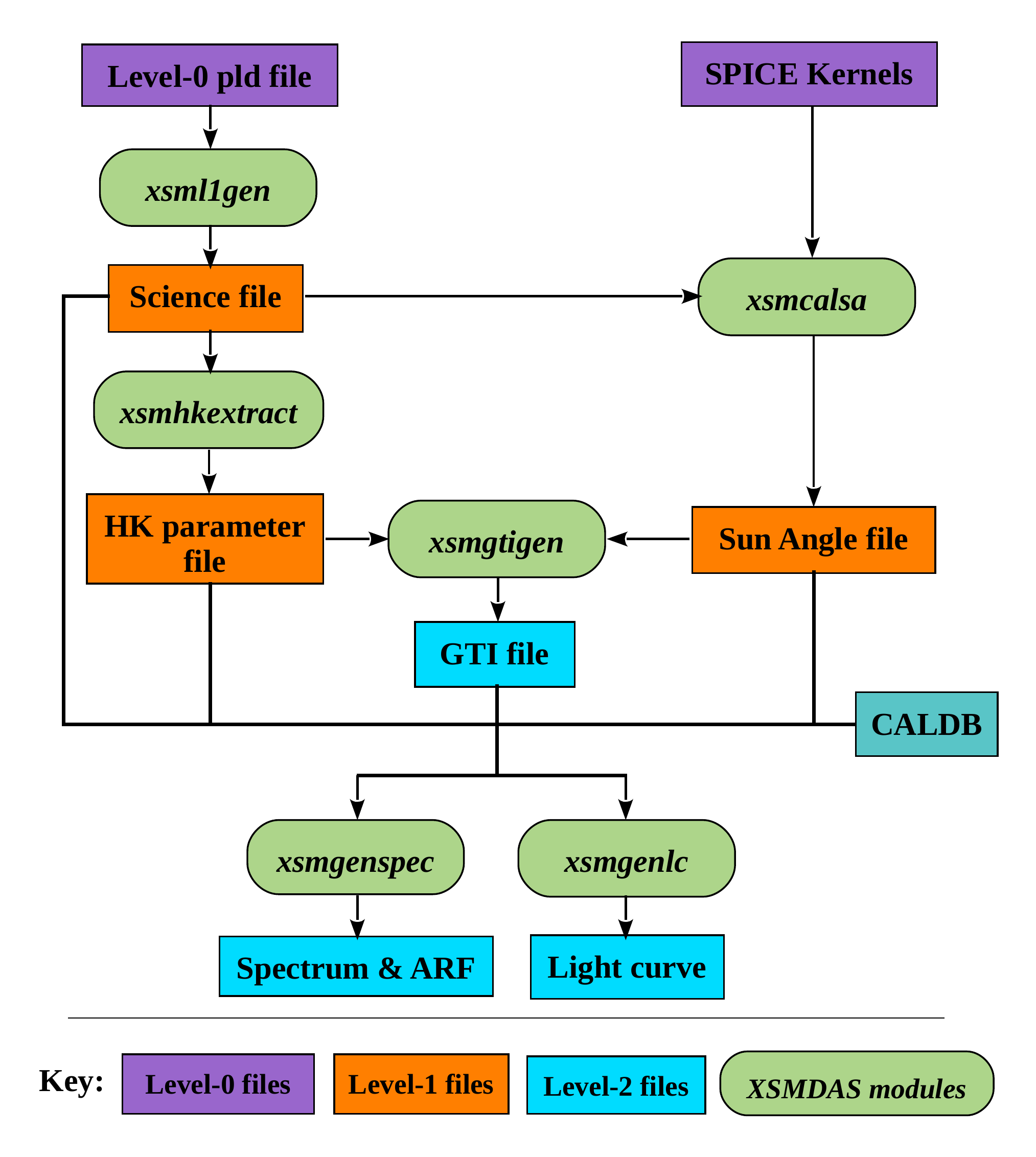}}
\caption{Work flow of the XSM Data Analysis Software (XSMDAS) showing its modules and 
their input and output files.}
\label{xsmdas_flowchart}
\end{figure}

Figure~\ref{xsmdas_flowchart} shows the flowchart of the XSMDAS where 
individual modules and their input and output files are shown.
The modules that generate level-1 files do not have any user-configurable
parameters and also do not use any calibration information. Hence, these are
executed only at the POC to generate level-1 data files, which are to be made available
to the users. On the other hand, the modules for generating level-2 science products require
various user inputs as well as the calibration data and hence the users may 
need to execute these modules with specific inputs to meet their requirements.

The XSMDAS is developed in C\texttt{++} and uses several standard libraries as well as few
third-party libraries. It uses the CFITSIO\footnote{\url{https://heasarc.gsfc.nasa.gov/fitsio/}} 
library for reading, writing, and manipulating
FITS files and PIL\footnote{\url{https://heasarc.gsfc.nasa.gov/lheasoft/headas/pil/pil.html}} 
is used for parsing user inputs.
PIL based input parsing allows users to provide inputs in an interactive manner (see section~\ref{xsmgenlc}) 
or as command-line arguments.
Specific modules also use the CSPICE\footnote{\url{https://naif.jpl.nasa.gov/naif/toolkit.html}} 
library for SPICE~\citep{acton96} data and 
\emph{TinyXML}\footnote{\url{https://sourceforge.net/projects/tinyxml/}} library for XML
file manipulation, as described in the subsequent section. The software can be compiled
on several Linux distributions and OS-X. It is tested
extensively to ensure proper execution, the correctness of products,
and exception handling. Further, the codes were tested for memory leaks using the Valgrind\footnote{\url{https://valgrind.org/}}
tool and the identified issues were resolved. The XSMDAS 
will be made available publicly, along with the first release of data sets 
from ISDA. Subsequent revisions of the software, if necessary, will also 
be provided from time to time.

\subsection{XSMDAS modules and algorithm}

XSMDAS is composed of eight modules; three each for the generation of level-1 and level-2
data products and two additional modules for adding spectra and generating
XML labels corresponding to the data product files.  A brief description of the
algorithm employed and the available options for all the
modules are given below. A complete list of user input parameters is given in the user guide
distributed along with the software. Help files can also be accessed with
the \emph{xsmhelp} command.

\subsubsection{xsml1gen}

This module ingests the level-0 binary data (.pld file) and generates the level-1
science data file in FITS format. It has options to either read a single level-0 file
and generate the respective level-1 file or to read multiple level-0 files and
write a science file with packets that correspond to observations of a single day
removing any duplicates.
\emph{xsml1gen} identifies valid data packets from the binary data, decodes the
UTC and DH time from the headers, and for each unique packet writes out the details such as
UTC, Mission Elapsed Time (MET) defined as the number of seconds from a reference
time (MJD 57754), DH time, and the XSM 2048 byte long data frame stripping the DH and level-0 headers
into different columns of the output FITS file. It also provides other useful information like
the fraction of good packets in each level-0 data set. It may be noted that this module merges 
the data from multiple level-0 files into day-wise level-1 files.

\subsubsection{xsmhkextract}

\begin{figure}[h]
\centerline{\includegraphics[width=0.99\columnwidth]{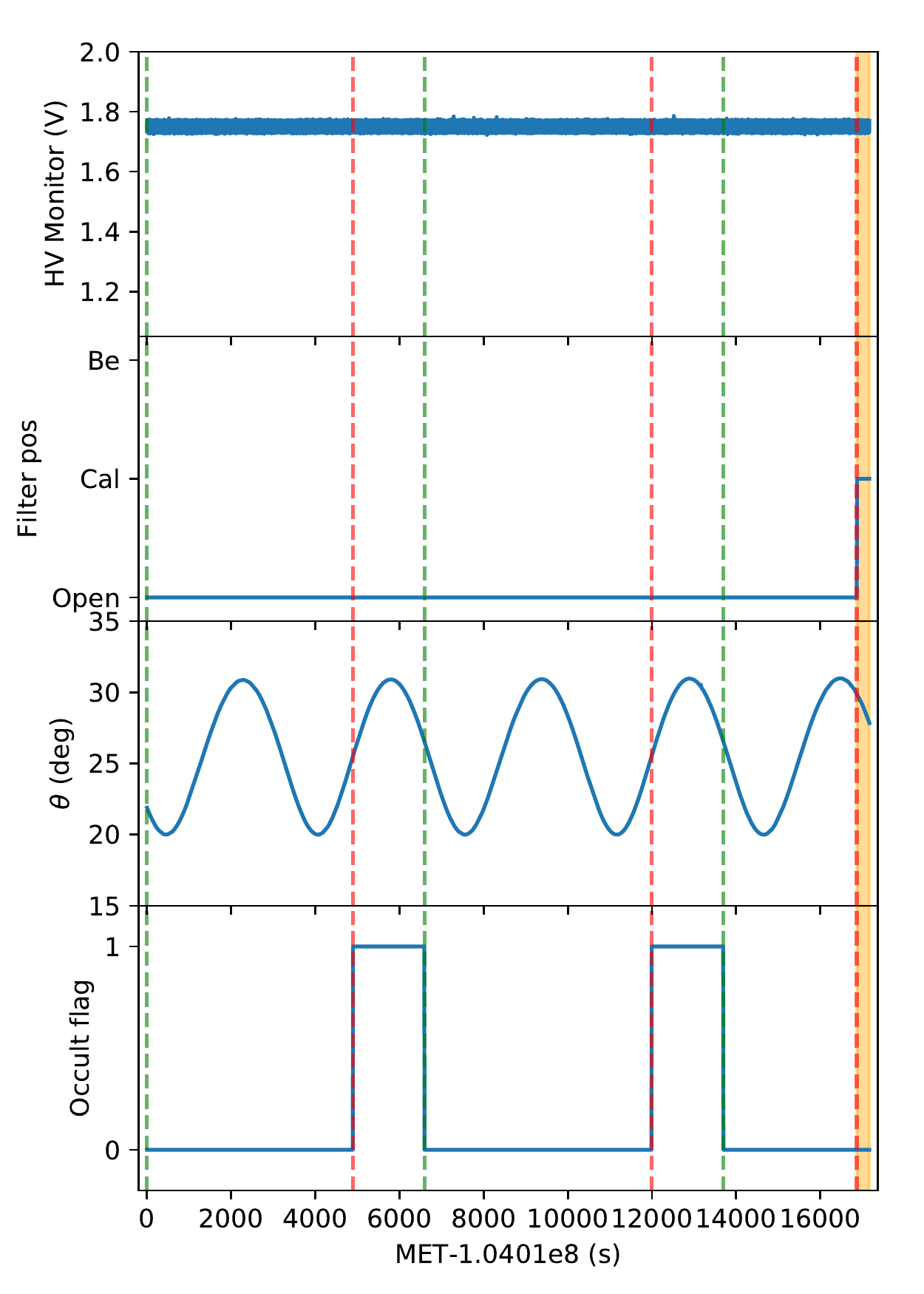}}
\caption{Two of the housekeeping parameters (upper two panels) and two solar observing geometry parameters (lower two panels) 
of the XSM are shown for a duration of $\sim$2.5 orbits. 
The vertical dashed lines mark the start (green) and end (red) of GTIs for solar observations whereas the 
shaded (orange) duration represent the GTI for calibration observation. 
The parameters are plotted from the output files of \emph{xsmhkextract} and \emph{xsmcalsa} and 
the GTIs are computed by \emph{xsmgtigen}.}
\label{xsm_hksagti}
\end{figure}

XSM data packets include housekeeping parameters such as various voltage levels, detector
temperature, and high voltage monitor. It also includes set values for various configurations
of the instrument, such as its low energy threshold. These parameters are required for
identifying good time intervals and obtaining appropriate calibration parameters
for a specific observation duration. They also serve the purpose of monitoring the
health of the instrument. \emph{xsmhkextract} reads the level-1 science file
generated by \emph{xsml1gen} and decodes these parameters from the data packets. The analog
measurements like voltages are recorded in the data as the respective ADC value, and to obtain the actual value; appropriate conversion factors are applied. For each data packet,
the decoded parameters are written into a row of the output HK FITS file, which will serve
as an input for level-2 product generation. Two of the HK parameters from the output file, 
HV monitor and filter wheel position, during an observation are plotted in the 
upper two panels of figure~\ref{xsm_hksagti}.

\subsubsection{xsmcalsa}

As the angle subtended by the Sun and hence the effective area varies during the XSM observation,
it is required to know the Sun angle in order to calibrate the light curves
and spectra. The projected position of the Sun in the detector is also required for calibration,
as it determines the gain factors to be applied for conversion of the PHA channels to energy~\citep{mithun20}.
This module computes such parameters that define the observational geometry, required
for further analysis of the data.
This computation requires determining the Sun's
position in Inertial Coordinate Reference System (ICRS) and then performing coordinate
transformations to the satellite's Body Coordinate System (BCS) and then to
XSM's Detector Coordinate System (DCS). For this purpose, \textit{xsmcalsa} uses SPICE
library routines~\citep{acton96}.

Input data required for these computations are provided in the form of SPICE kernels.
The spacecraft position, attitude, and clock information in the form of SPICE
SPK, CK, and SCLK kernels are part of the level-0 data set. Ephemeris of solar system
objects is also available as standard SPK files.
The other kernels required are a frame kernel that provides the Euler angles between the spacecraft
BCS and XSM DCS and an instrument kernel that defines the field of view of the XSM.
Both these kernels are generated in required formats based on ground measurements.
\textit{xsmcalsa} loads all these kernels and first verifies whether the orbit and
attitude information span the observation duration of the science data. Then,
appropriate functions from SPICE API are used to perform the required coordinate
transformations to obtain the Sun's position vector in XSM DCS for each second.
From this, polar Sun angle ($\theta$) defined as the angle from the instrument boresight, and
the azimuthal Sun angle ($\phi$) are computed. It also checks whether the Sun is
occulted by the Moon and whether it is outside the FOV of XSM and assigns values 
to the respective flags.
Using the polar Sun angle and geometry of the detector and collimator, the projected
position in the detector is calculated. All these parameters as a function of time
are written to an output FITS file. As an example, the variation of polar Sun angle 
($\theta$) and the occultation flag during an observation is shown in the 
lower two panels of figure~\ref{xsm_hksagti}. 

\subsubsection{xsmgtigen}

The science data products are to be generated by selecting data during Good Time Intervals (GTI)
when various housekeeping and geometry parameters are within their favorable limits. This module named \textit{xsmgtigen}
computes these GTIs. The required parameters are read from the HK and sun angle files generated by the
previous two modules of the XSMDAS, and their limits are read from a filter file provided by the user.
It also provides an option for the user to provide additional GTI as another input file.
The module first computes the basic GTI, which corresponds to the duration when the data
is available; if contiguous data is available, this GTI will have only one interval
for the entire observation duration. Then, the GTI where all the parameters
are within the defined limits is determined and its  intersection with the basic
GTI and the user-defined GTI (if provided) are computed to get the final GTI, which
is then written into a FITS file.

The default run of \textit{xsmgtigen} at the POC generates a GTI when the Sun observations are
available (within FOV and not occulted) and useful based on the housekeeping parameters
of the instrument. In the example given in figure~\ref{xsm_hksagti}, the vertical dashed 
lines show the default GTIs for solar observations which include durations when 
the Sun is not occulted and within the instrument FOV and the filter wheel is 
in open position. The users are expected to apply this minimum set of GTI selection
criteria in their analysis, which is defined in the standard filter file provided along with
the software. Additional constraints on parameters can be included based on
specific requirements.
Further selections of time intervals not based on the HK or geometry parameters
for which the spectrum  or light curve are to be generated can be done
by providing the user GTI input.
This will be convenient if the user wishes to select multiple intervals of data
to analyze, for example, based on the observation duration of another instrument.
Apart from selecting the time ranges for solar observations, this module also can be
used to generate time intervals when the XSM is acquiring calibration data
by selecting the filter wheel position to be calibration source instead of open or
to obtain intervals when the XSM is acquiring background data with Sun out of the
complete FOV or occulted by the Moon. The duration marked by shaded region in figure~\ref{xsm_hksagti} 
shows the GTI for calibration. 

\subsubsection{xsmgenspec}

As the name suggests, this module generates spectra from XSM data based on various
user input parameters. It requires the outputs from all the previous modules and
provides time-series or time-integrated spectrum in standard PHA file
formats (type-I or type-II)\footnote{\url{https://heasarc.gsfc.nasa.gov/docs/heasarc/ofwg/docs/spectra/ogip_92_007/node5.html}}
compatible with the X-ray spectral analysis tools like XSPEC~\citep{arnaud96} and
ISIS~\citep{2000ASPC..216..591H}. It also provides the corresponding ancillary response file (ARF)
required for the analysis of the spectrum.

\begin{figure}
\centerline{\includegraphics[width=0.99\columnwidth]{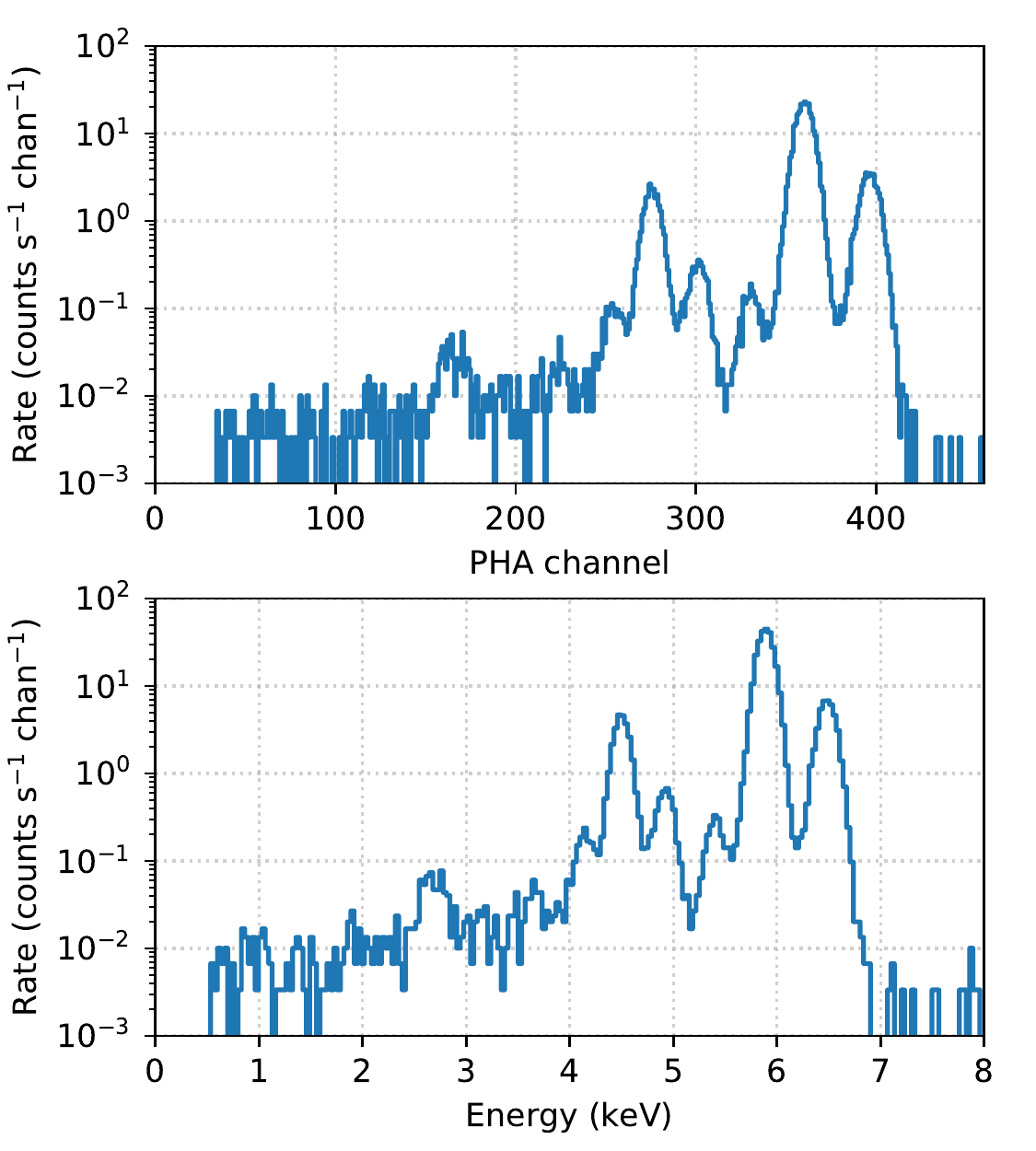}}
\caption{XSM calibration source spectrum in PHA (top) and PI (bottom) channels. PHA channels have 
1024 bins with a bin size of $\sim$16.4 eV/channel, which varies with the instrument gain. The PI channels are 
512 bins with 33 eV binsize starting from 0.5 keV.}
\label{xsm_spec}
\end{figure}

\textit{xsmgenspec} selects all data frames within the GTI duration and adds together
the spectra as required. In the case of time-integrated spectrum, all spectra within
the user-provided start and end time are added together. If the user opts for a time-series
spectrum, data between the start and end times are binned to the user-provided time
bin size. While adding together the spectra from each raw data frame, it obtains
the gain factors from the calibration database corresponding to the instantaneous observing
conditions and then resamples the PHA spectrum to Pulse Invariant (PI) channels
that correspond to 512 energy bins from 0.5 keV with a bin size of 33 eV.
For each count detected in a PHA channel in a one-second interval,
a uniform random number in the (0,1) range is generated and added together with the integer PHA value. This floating-point
PHA value of the event is then multiplied with gain factors to get the energy, and from the
energy, the respective PI channel is identified and the count is added to that channel.
Figure~\ref{xsm_spec} shows the raw PHA spectrum and the resampled 512 channel PI spectrum for a calibration observation. It can be seen that the line energies are correctly determined in the PI spectrum.
Statistical and systematic uncertainties on the counts in each channel are computed based on
Poisson statistics and from the calibration database, respectively. These are
also recorded in the output spectrum file.

For the duration of the spectrum, the average effective area is also computed
based on the Sun angle and the calibration database information, and this is
written out as an ARF file. As the effective area variations
are accounted in the ARF file, the counts in the spectrum are not corrected for the area
and hence do not directly provide an estimate of the incident flux. The spectral analysis
would also require a redistribution matrix file (RMF) along with the spectrum and ARF.
The name of the appropriate RMF file is written in the header of the spectrum file, and the RMF files are distributed as a part of the calibration database. 

Instead of generating a separate ARF for each duration of the spectrum, \textit{xsmgenspec} also 
provides an option to generate spectra corrected for the effective area variation. 
In this case, the spectra will be scaled such that they represent counts detected by the XSM 
when the Sun is on-axis, and the header of the file includes the respective response file (.rsp) 
which contain the redistribution matrix and on-axis effective area. 
The standard spectrum file provided as part of the data distribution a time-series 
spectrum with a time bin size of one second falls into this category and it 
is meant for spectral analysis using OSPEX as discussed in section~\ref{sec:dataproduct}.

Apart from generating the calibrated science products, \textit{xsmgenspec} also has options
to create outputs without applying calibration. This can be done by selecting the available 
appropriate user input options and is particularly useful for monitoring the
instrument performance at the POC. 

\subsubsection{xsmgenlc}
\label{xsmgenlc}
Light curves, which are time series of counts integrated over a given energy range, 
are created by the \textit{xsmgenlc}. For each second data of the XSM, this module adds together
the counts within the user-specified energy range after correcting the raw one-second
spectral data for gain and the ratio of instantaneous effective area to that for the on-axis case.
It is then binned to the desired time interval, which can be integer seconds, and
is written out in a standard FITS file format\footnote{\url{https://heasarc.gsfc.nasa.gov/docs/heasarc/ofwg/docs/rates/ogip_93_003/ogip_93_003.html}}.
As the effective area variations are corrected, the light curves provide the rates when
observed on-axis with the XSM. The errors on the count rates are computed by considering Poisson statistics 
for raw counts and then propagating them and are written into the respective 
column in the output file. Standard processing at the POC generates a light curve 
for the entire energy range of 1--15 keV with a bin size of one second and is included as a part of the data archive.

\begin{figure}
\centerline{\includegraphics[width=0.99\columnwidth]{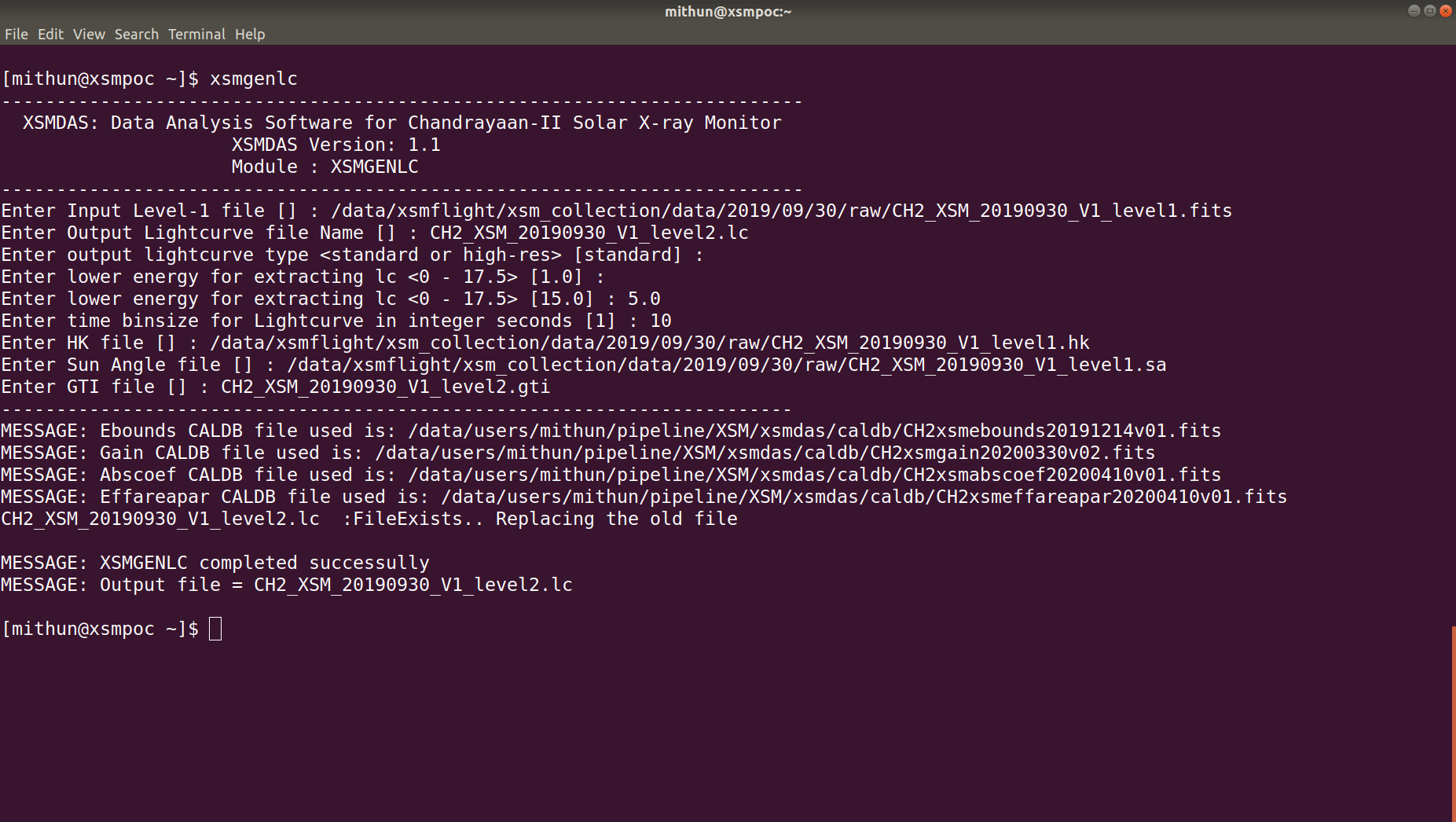}}
\caption{Interactive command-line user interface for \textit{xsmgenlc} module of 
the XSMDAS using PIL.}
\label{uixsmgenlc}
\end{figure}

As an example for the user interface of the XSMDAS modules, figure~\ref{uixsmgenlc} shows 
the screen shot of the interactive execution of \textit{xsmgenlc}. Here, the user has selected 
to generate a light curve for 1--5 keV with a time bin size of 10 seconds. 
Apart from the option to generate light curves for any energy range with cadence above one
second, there is also provision to generate light curves with a time bin size in multiple of 100 ms
for the three pre-defined energy ranges. In this case, instead of decoding the spectral
data from the XSM frames, the high time resolution light curves for three channels from the
header are used. The light curves for the three energy ranges are written out as three
extensions of the output FITS file.

\subsubsection{xsmpds4gen}

As the XSM data is planned to be archived following the PDS4 standards, each of the files
generated by the above modules needs to have an associated XML label file defining the
contents of the FITS data files. \emph{xsmpds4gen} generates these PDS4 XML labels
associated with all level-1 and level-2 FITS files. For each type of data product,
a template XML file is made where all mandatory classes and attributes for the
PDS4 compliance are included according to PDS4 information model version 1.11\footnote{\url{https://pds.nasa.gov/datastandards/documents/im/v1/index_1B00.html}}.
The actual values of attributes that have fixed values are included in the
template, whereas the others have placeholder values.
The \emph{xsmpds4gen} reads the FITS product file and generates the respective XML
label using the appropriate template XML file. The attributes that had placeholder
values are replaced with the actual values based on the contents of the FITS file.
A simple open source C\texttt{++} XML parser named \emph{TinyXML}
is used for manipulation of XML files. 
The XML labels generated by this module are verified
using the PDS4 validate tool\footnote{\url{https://pds.nasa.gov/tools/about/pds4-tools/v1/index-1B00.shtml}}.

\subsubsection{xsmaddspec}

This tool provides the facility to add together multiple spectrum files generated by
\emph{xsmgenspec} to create a single spectrum output. It is particularly useful when
there is a need to combine the XSM spectra from multiple days of observation.
The module takes care of the propagation of statistical and systematic errors of the
channel-wise counts while adding the spectra together. It also combines the ancillary
response files that correspond to the individual spectra and provide an output ARF file
to be used with the added spectrum.

\subsection{Calibration database for XSM}

Various calibration parameters such as gain, spectral redistribution, and effective area,
which are required for the analysis of XSM data, are derived from extensive
ground calibration experiments~\citep{mithun20}.
These data are stored in a set of FITS files that form the calibration database (CALDB)
of XSM, which is accessed by the XSMDAS modules to generate the products.
The XSM CALDB design follows a philosophy similar to the HEASARC's CALDB
system\footnote{\url{https://heasarc.gsfc.nasa.gov/docs/heasarc/caldb/caldb_intro.html}} meant for 
X-ray astronomy missions.
Each aspect of calibration data is stored in individual FITS file BINTABLE
extensions with appropriate structure. In case of any update in a particular
calibration data, a new file with the same format as the previous one is
included in the new release while retaining the old versions. A calibration
index file (FITS format) included in each CALDB release keeps a record of
the file names with the latest calibration data of each type and includes the
names of older versions of the same data with their quality flag marked as bad. 
The index file also defines the validity start and end time for each file, thus allowing
multiple good files of the same type valid over a different range of times.
This is particularly useful if some of the calibration parameters (e.g., gain)
varies over the mission duration and different set of values are to be used
for observations at different times.

With such a design of the CALDB system, the calibration file names are not hard-coded in the
software. XSMDAS accesses the index file of the CALDB and identifies the appropriate
calibration data file based on the observation time of the data being processed.
This allows updates of the CALDB without any updates in the data analysis software
unless there is a change in format or addition of new types of calibration files.
The XSM CALDB includes six types of calibration files other than the calibration
index file, the contents of which are briefly described below. See \cite{mithun20} 
for the details on how these parameters are estimated.

\begin{enumerate}[leftmargin=*,label=(\roman*)]
\item Gain: This file includes the gain and offset of the XSM
over a grid of TEC current and interaction position in the detector.
\item Effareapar: This file includes the parameters required
for the computation of effective area, defined over an x-y grid of
position of the Sun in the field of view of XSM.
\item ARF: Ancillary Response File includes the effective area
as a function of incident energy over a grid of Sun positions as in the Effareapar file.
\item RMF: Redistribution Matrix File defining the spectral redistribution
function in Pulse Invariant (PI) channels over a grid of incident energies.
\item Syserror: Channel-wise systematic errors to be included in the
spectrum are recorded in this file.
\item Ebounds: This file defines the nominal energy range for the PI
channel definition.
\item Abscoef: This file includes the X-ray absorption coefficients of materials
that form the entrance window (Be), dead layer ($\text{SiO}_2$), and the detector (Si) 
obtained from NIST.
\end{enumerate}

The XSM CALDB will be made available along with XSMDAS during the data release. 
This would also include the refinement of various calibration parameters from 
the in-flight observations, results of which will be reported elsewhere.
Further updates in the calibration if any would also be released as 
subsequent versions of the CALDB.

\section{XSMQLD: XSM Quick Look Display}
\label{sec:xsmqld}

 \begin{figure}
 \centerline{\includegraphics[width=1.0\columnwidth]{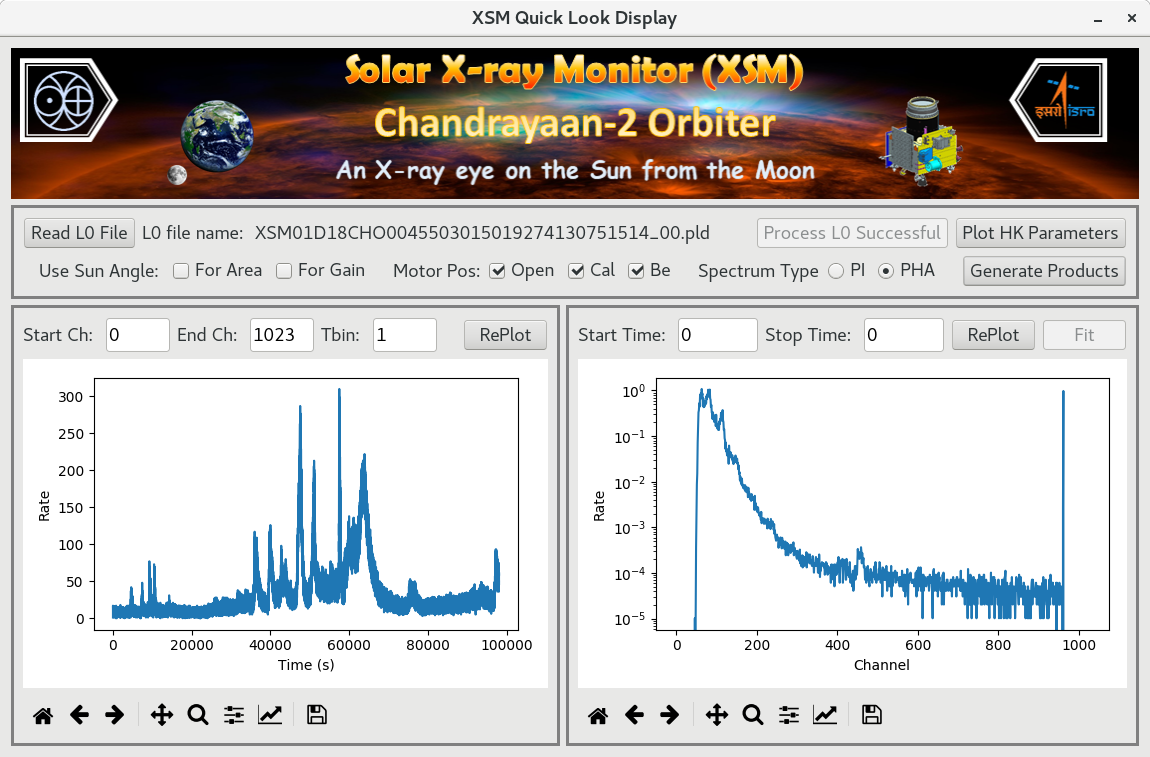}}
 \caption{XSMQLD window displaying the data during 30 September to 01 October 2019, 
 when XSM recorded its first B class solar flare. The plots in the window are the un-calibrated 
 light curve and spectrum.}
 \label{xsmqld_sun}
 \end{figure}

The XSM Quick Look Display (XSMQLD) provides visualization of the XSM data
products without any calibration and that of the housekeeping parameters, 
using the level-0 binary data as the input.
The XSMQLD is designed as a graphical front-end that uses XSMDAS modules to carry out
the processing in the back-end. It is developed in python, making use of PyQt4,
the python binding for Qt GUI toolkit. Figure~\ref{xsmqld_sun} shows the XSMQLD
window with light curve and spectrum during one of the observations. As seen
from the figure, the main window includes plots of the basic products of XSM,
the light curve, and the spectrum along with other widgets for user inputs 
to view specific products.

XSMQLD allows the user to select a level-0 file that needs to be visualized.
It executes the modules \emph{xsml1gen} and \emph{xsmhkextract} and generates
level-1 files and further generates the level-2 products executing the relevant
XSMDAS modules. The default generated light curve and spectrum without applying
any calibration are displayed in the QLD plot windows. The range of plots
can be adjusted with the options available. The XSMQLD also has provisions for plotting
light curves and spectrum for specific filter wheel positions such as calibration.
It also allows us to plot the light curve for a specific energy range or spectrum for
a specific time range. All these specific selections are carried out by executing the
XSMDAS modules with the inputs provided by the user in the GUI window. There are 
also options in the software to view the plots of various housekeeping parameters decoded 
from the data.

 \begin{figure}
 \centerline{\includegraphics[width=1.0\columnwidth]{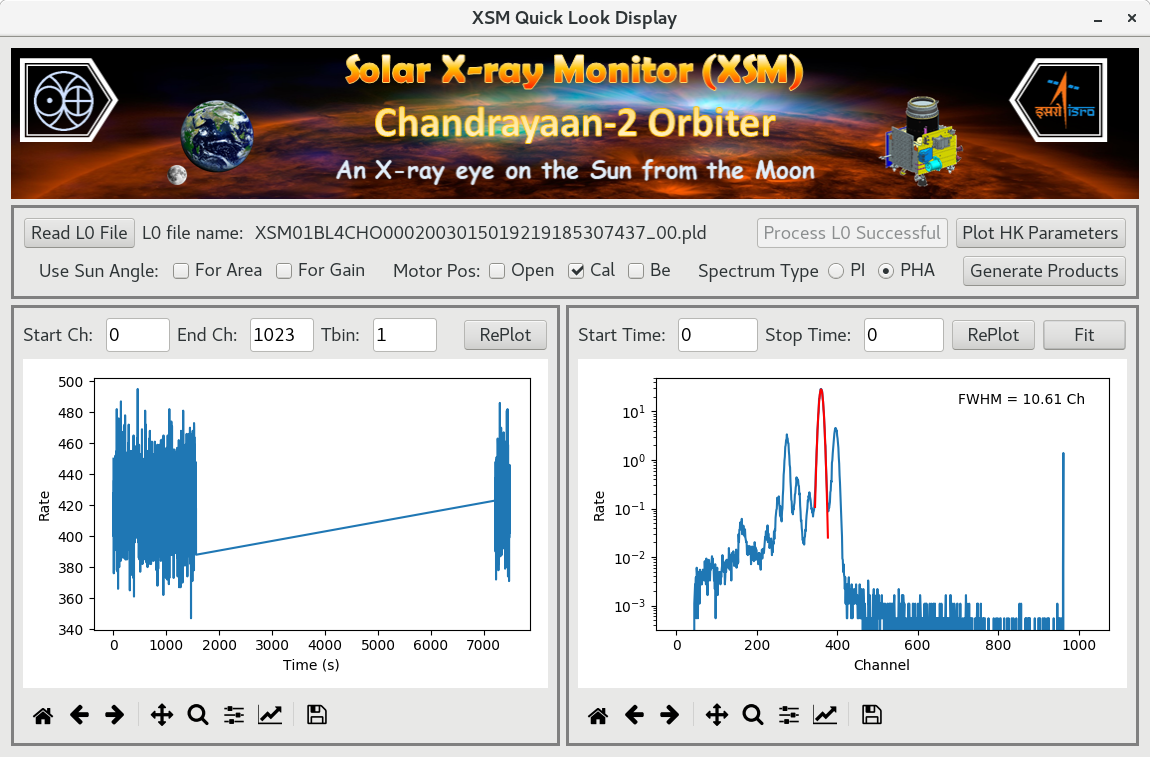}}
 \caption{XSMQLD window showing the calibration data acquired on 08 August 2019 
 when it was powered on the first time in space. The spectral line is fitted to 
 determine the energy resolution.}
 \label{xsmqld_cal}
 \end{figure}

Spectral resolution measured by the Full-Width Half Maximum of a spectral line
from the calibration source determines the performance of XSM. Hence, the XSMQLD
also has provisions for obtaining this parameter by fitting the 
calibration source spectrum. Figure~\ref{xsmqld_cal} shows the QLD window with 
the calibration source spectrum, where the line at 5.9 keV is fitted with a 
Gaussian to provide the measurement of the spectral resolution. 
The XSMQLD was particularly useful during the commissioning phase of the 
instrument in space and was used extensively during that period to quickly 
examine the data once it is downloaded from the spacecraft so that further 
operations could be planned. 

\section{XSMDMS: XSM Data Management-Monitoring System}
\label{sec:xsmpocms}

After the spacecraft operations were regularized, XSM has been operating almost
continuously, acquiring data with only brief durations of power off for orbit
maneuvers and other specific mission operations. The data downloaded at
regular intervals undergo ground pre-processing at ISSDC and the level-0
data sets are made available for the POC. Higher levels of processing using
the XSMDAS modules is carried out at the POC, which is managed by
the XSM Data Management-Monitoring System (XSMDMS).

The XSMDMS is a collection of software tools to carry out the automated
level-1,2 data processing using XSMDAS modules, maintain an SQLite database
recording the relevant information of each data set, and update an
internal web application with the plots of house-keeping parameters, light curves, 
and spectra to facilitate monitoring of the instrument health and performance.
The overall workflow of XSMDMS, details of the database, and the web application are discussed here.

\subsection{XSMDMS work flow}

Data processing at the XSM POC fall into two categories: (i) the processing of the level-0
data sets from each download individually to generate respective higher-level products;
and (ii) the processing to generate day-wise level-1 and level-2 products combining all
the available level-0 data sets for the respective day. Products of the former are meant
for use at the POC alone for purposes like examining the data quality, whereas the
day-wise merged products will go into the XSM data archive. In both cases, apart
from generating calibrated level-2 products (spectrum and light curve) selecting the duration
of solar observations, uncalibrated spectra and light curves for the entire
observation are also generated for monitoring instrument background and other
aspects. Whenever calibration source observations are available in the data,
spectrum for that duration is also separately extracted during POC processing
to aid monitoring of performance and gain of the spectrometer.

 \begin{figure}
 \centerline{\includegraphics[width=1.0\columnwidth]{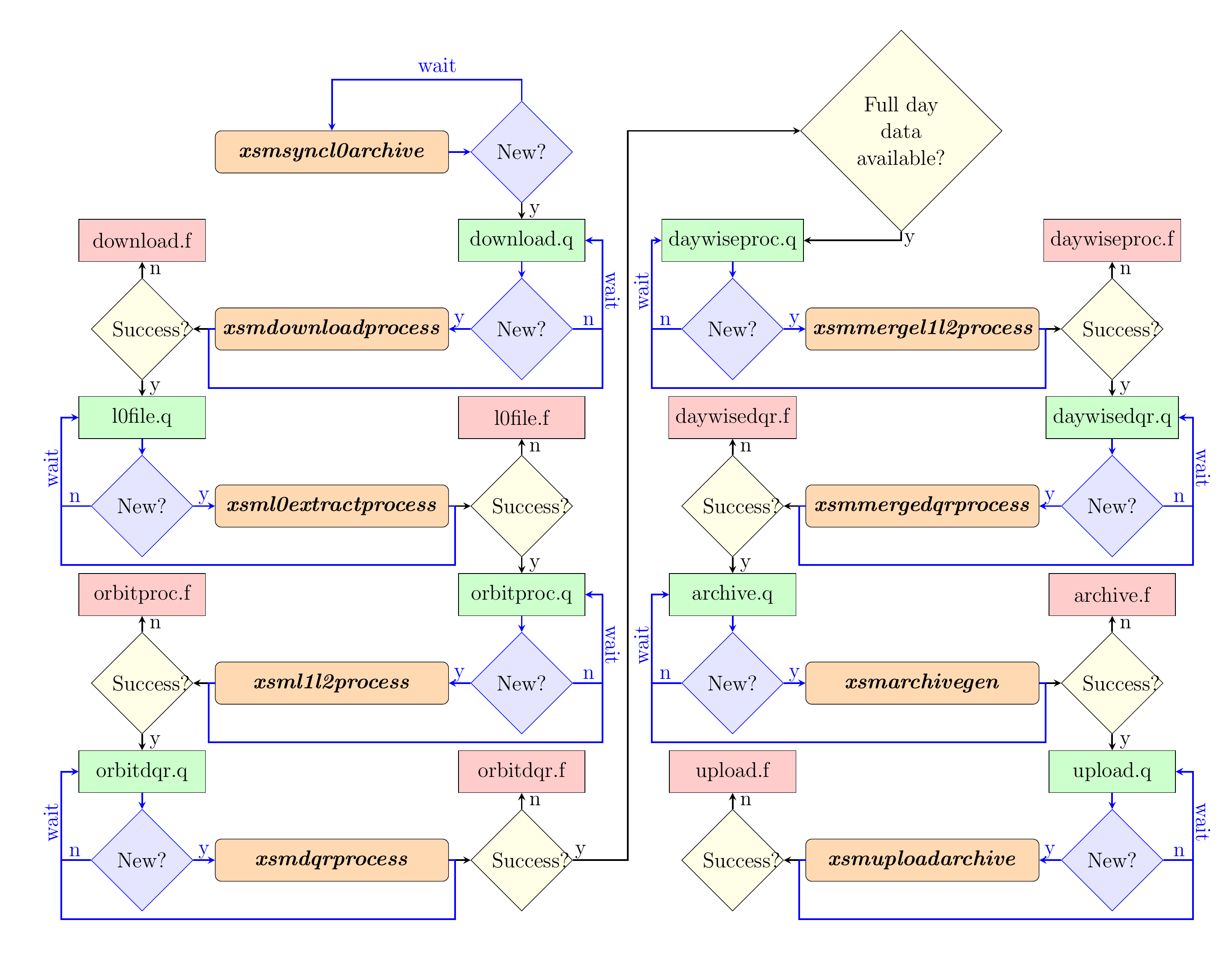}}
 \caption{Flowchart of XSMDMS. The chain on the left correspond to the individual 
 processing of level-0 data sets, whereas that on the right carries out day-wise 
 merged processing. Input queues are named *.q and the *.f files list the failed processes.
 Each of the closed loops in the diagram shown in blue color runs perpetually with 
 pre-defined wait periods.  
 }
 \label{poc_proc}
 \end{figure}

XSMDMS is designed as a collection of scripts that are executed in a pipeline mode
and is responsible for carrying out all these data processing activities at
the POC. Figure~\ref{poc_proc} shows the flow chart of the
complete processing chain at the POC, where the modules on the left are meant for individual level-0 data set processing, and those on the right are for generation of day-wise files. XSMDMS also updates an SQLite database with the information of
each individual and day-wise data set and creates web pages with plots summarising
each observation.

Individual modules or processes of the XSMDMS shown in figure~\ref{poc_proc} are bash scripts
which internally call XSMDAS tasks for level-1,2 processing and a set of python
scripts for updating the database and creation of web pages. The modules are divided
into logical blocks with each having a specific task and they are linked
together as a processing chain with the help of a simple text-based queue system.
All of them, except the first one, have an associated input queue, which is nothing but a plain
text file. Each module periodically checks the input queue file to find the next data set to
be processed, and after its successful execution, respective entry is
made in the subsequent module's input queue while removing the particular
entry from its input queue. In the case of a failure in a module for one data set,
it will not be added to the next process's queue; instead, it is recorded in the
respective module's failure queue with reference to appropriate log files
to facilitate manual verification, if required. This system also allows
the re-execution of processes starting from any point in the chain for 
only certain data sets, if such a requirement arises.
A brief description of the functionality of each module of the 
XSMDMS is given below.    

\begin{itemize}[leftmargin=*]
\item \textit{xsmsyncl0archive}: Level-0 data sets in the form of compressed files 
for each orbit's download are generated at the ISSDC server at the end of level-0 processing. 
This module periodically checks the level-0 archive at ISSDC and compares it with the 
data sets already present at the POC and identifies if any new data is available. 
If any new level-0 data sets are present, they are added to the next module's input queue.
\item \textit{xsmdownloadprocess}: This module downloads the new level-0 tar files
from ISSDC to POC, and checks whether the download is complete by verifying the
downloaded file size with that recorded in an associated trigger file. In the case of
any mismatch in the file size, it attempts the download twice more before
recording it as a failed download. If a level-0 tar file is downloaded successfully,
it is added to the following module's input queue.
\item \textit{xsml0extractprocess}: It extracts the contents of level-0 tar file
to designated directories after verifying the existence of all required files within it.
Sometimes each tar file may contain multiple segments
of payload data recorded in different binary files and each payload data file (.pld)
having associated auxiliary files. In such cases, they are segregated and extracted
to individual directories and are added as separate entries to the successive
process queue.
\item \textit{xsml1l2process}: This module carries out the processing of each level-0
data set to generate respective level-1 and level-2 products. This is done by executing
individual modules of the XSMDAS with appropriate inputs. The level-2 data product
generation is done for durations when the Sun is observed applying calibration,
the entire observation duration without any calibration,
and durations when the calibration source spectrum was acquired, depending on
whether these conditions are met in a particular data set.
\item \textit{xsmdqrprocess}: This module is responsible for the execution of python scripts to generate web pages with a summary of each observation and to update the database discussed in subsequent sections. It also triggers the day-wise merged 
processing if all the data for a given day of observation are available, which 
is identified by querying the database.
\item \textit{xsmmergel1l2process}: This module has functionality similar to
\textit{xsml1l2process}, except that it carries out higher-level processing
to generated day-wise product files. It queries the database using a python
function and identifies all level-0 data sets available for a given day of
observation, and these are provided as inputs to XSMDAS modules. Like in
\textit{xsml1l2process}, three types of level-2 products are generated
based on availability.
For the raw and calibrated data products, respective
PDS4 labels are also generated by executing \textit{xsmpds4gen} with appropriate
inputs.
\item \textit{xsmmergedqrprocess}: Its functionality is similar to \textit{xsmdqrprocess},
and it creates summary web pages for day-wise products and update the database.
\item \textit{xsmarchivegen}: This carries out validation of the day-wise products that
are to be included in the archive and bundles them together into a compressed file.
It also generates an XML label associated with the day-wise archive file that includes
all relevant metadata required by the data dissemination application.
\item \textit{xsmuploadarchive}: This last module uploads the XSM day-wise data
archive back to ISSDC for dissemination.
\end{itemize}

\subsection{SQLite database}

The XSMDMS maintains an SQLite\footnote{\url{https://www.sqlite.org/index.html}} database
for storing the information of the XSM data sets and their processing status. The database
includes two tables corresponding to the individual level-0 data set processing (\textit{xsmorbitdata})
and day-wise merged processing (\textit{xsmdaywisedata}).
Both tables include columns that provide basic information such as
start and times of observation, exposure times, processing status, etc. In the case of
individual orbit data processing, the level-0 file is the primary key, whereas, for the
day-wise processing, the day of observation is the primary key.
As an example, table~\ref{dbase_columns} lists the columns of the
database table \textit{xsmdaywisedata}.

 \begin{table}
 \caption{List of columns in the database table \textit{xsmdaywisedata}. In the case 
 of status flags, the remarks mention the criteria when the the flag is set to true.}
 \label{dbase_columns}
 \begin{tabular}{l l}
 \hline
 \hline
 Column name & Remarks \\
 \hline
 DayOfObs  &  Primary key (yyyymmdd) \\
 StartTime  &  UTC string\\
 StopTime   &  UTC string\\
 Tstart   &  MET (s)\\
 Tstop   &  MET (s)\\
 TotalExposure  & seconds\\
 SunExposure   & seconds\\
 procstatus & Processing successful \\
 calstatus & Calibration data present \\
 sunstatus & Sun data present \\
 archivestatus & Archive generation successful \\
 uploadstatus & Data uploaded \\
 indexfile  & HTML file observation summary\\
 \hline
 \hline
 \end{tabular}
 \end{table}

In order to access and update the database, the python module
sqlite3\footnote{\url{https://docs.python.org/3/library/sqlite3.html}}
that provides an SQL interface is used. The \textit{xsmdqrprocess} and
\textit{xsmmergedqrprocess} modules of XSMDMS executes python scripts
that update the respective database tables. The required information on
each data set is read from the FITS file headers, and entry is made in
the database accordingly. In the case of day-wise processing, the further modules
for archive generation and upload will also update the respective entries
in the database.

The orbit-wise database table is also used by the module \textit{xsmmergel1l2process}
to identify level-0 files that contain data for a particular day.
This selection excludes any level-0 data set which has failed in its individual
processing so that only validated level-0 data sets are used for generating
the day-wise archive products. The database also makes it easy to identify
any duration where the data is missing so that it can be conveyed to mission
operations and ground pre-processing teams for necessary action. It is
also used to maintain a web application, as discussed in the next
section.

\subsection{Web application for observation summary}

Another component of the XSMDMS is a web application that provides a summary of 
each observation. It may be noted that the purpose of this web application is 
for monitoring the instrument operation and data and hence is hosted internally 
at the POC. 
For each of the orbit-wise and day-wise data set,
the summary web page includes basic information on the data, plots of the raw
light curve and spectrum, the solar light curve and spectrum, calibration spectrum
(if present), house-keeping parameters, observing geometry parameters, and
instrument settings for the given observation. This web page is a simple
static HTML file generated by python scripts that are executed by
\textit{xsmdqrprocess} or \textit{xsmmergedqrprocess}. 
These scripts read the FITS product files and generate plots, which are then included
in an HTML file. The HTML summary file name is then included in the
database entry for the respective data set.

The website's main page lists all orbit-wise or day-wise
observations obtained from the SQLite database with links to
the respective summary pages. Essential information like
start and end time of observation are also listed along with
the identifier to each observation. Summary of observations for
new data sets will become available as soon as the processing is
complete and respective entries are made in the database.
This web site listing observations is
implemented using the flask\footnote{\url{https://flask.palletsprojects.com/en/1.1.x/}}
web application framework in python, which allows rendering the contents
of SQLite database tables in a web application. The web application
is deployed with Apache on the POC server and is available only for
internal access.

 \begin{figure}
 \centerline{\includegraphics[width=1.0\columnwidth]{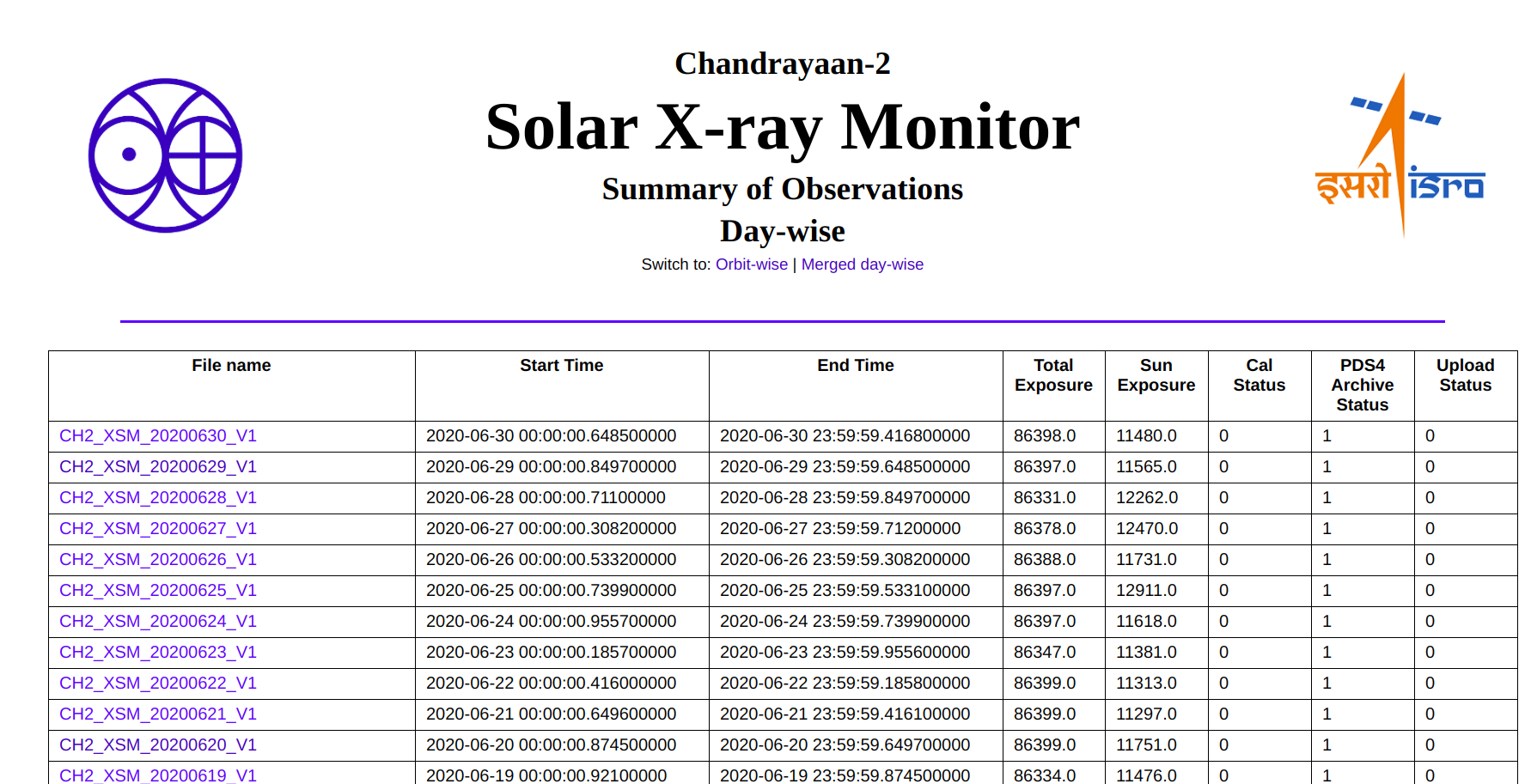}}
 \caption{Main web page of the XSMDMS web application listing the day-wise XSM observations. 
 Hyperlinks on the file names lead to individual observation summary web pages.}
 \label{webapp}
 \end{figure}

Figure~\ref{webapp} shows a screenshot of the web page with the list of 
day-wise observations and associated information. The hyperlinks on 
the file names lead to the summary page of the respective observation. 
The web application is frequently used by the XSM team to monitor the instrument parameters, to know the 
processing status, and more often to see the solar light curves and 
spectra from each observation.    

\section{Data archive and utilization}
\label{sec:dataproduct}

\begin{figure}
 \centerline{\includegraphics[width=0.95\columnwidth]{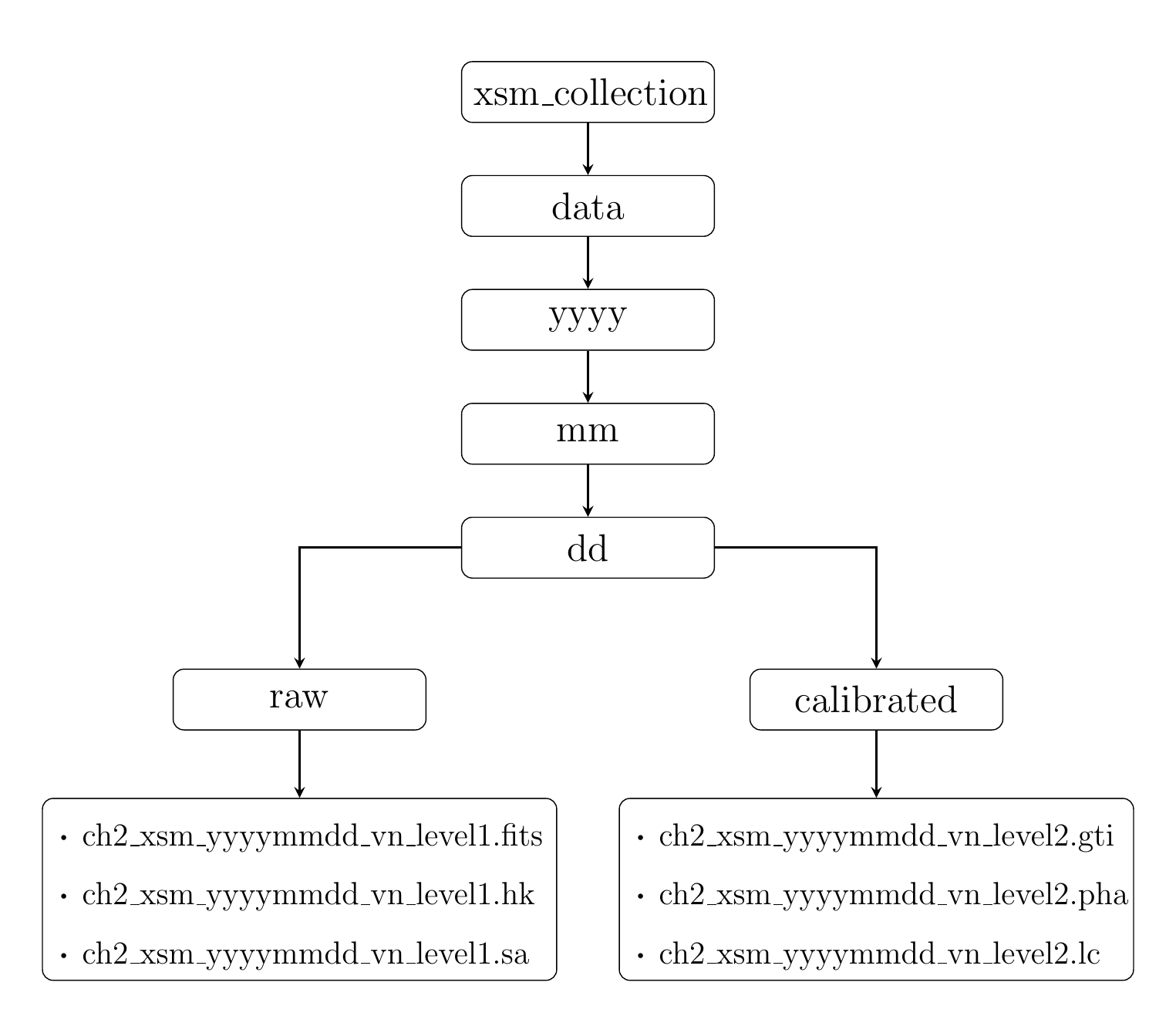}}
 \caption{XSM data archive structure showing the organization of data files. All data files 
 are in FITS format and each file will have an associated PDS4 XML label.}
 \label{archive}
 \end{figure}

The XSM data archive includes  `raw' (or level-1) data and `calibrated' (or level-2)
data created by the XSMDAS. All data files are in FITS format containing data corresponding
to the one-day duration and are organized into day-wise directories, as shown in the
archive structure in figure~\ref{archive}. The calibrated data sets are the outputs of default processing at POC with a standard set of parameters like time bin size or
energy range. However, it is possible to generate calibrated products with any desired
set of binning parameters from the raw data with the help of XSMDAS modules.

The daily data set of the XSM includes a maximum of six files, three of which are under the raw 
directory and three under the calibrated directory. 
If, on a particular day of observation, there were no periods when the XSM observed
the Sun, there will be no calibrated data for that day. The contents of the
six data files of the XSM are briefly described below.

\begin{enumerate}[leftmargin=*,label=(\roman*)]
\item  Science data file (*.fits): It tabulates the raw XSM data frames and the respective UTC,
Mission Elapsed Time (MET), DH clock time, and XSM clock time.
\item  HK parameter file (*.hk): Various housekeeping parameters and instrument settings decoded
from the data are provided as a function of time.
\item  Sun angle file (*.sa): This file includes various geometry parameters for the observation
duration, such as the polar and azimuthal angles of the Sun, the FOV and occult flags, 
and the projected radial position of the Sun on the XSM detector.
\item  GTI file (*.gti): This file provides the good time intervals from the default processing
that excludes durations when Sun is out of FOV or occulted, any of the instrument parameters
not within the desired range, and the times when data is not available. Each GTI is recorded
as a row in the two-column table listing start and end time of the interval in MET.
\item  Spectrum file (*.pha):  Standard spectrum file included in the XSM archive contains time-series
spectra (in type-II PHA format) with a time bin size of one second, where the counts are scaled 
to on-axis observations. Each row of the binary table records the start and end times, 
the spectrum, statistical and systematic errors.
\item  Light curve file (*.lc): The standard light curve provided is for a time bin size of 
one second integrating counts over the full energy range of XSM (excluding the ULD events). 
Count rate corrected for the instantaneous effective area as a function of time is tabulated in
this file.
\end{enumerate}

In the XSM archive, all the data files have an associated PDS4 XML label file with
the metadata. These XML labels include details such as a unique logical identifier to the
data set, observation details, the revision history of the data set,
and the structure of the respective FITS data file.
This allows tools like PDS4viewer\footnote{\url{http://sbndev.astro.umd.edu/wiki/PDS4_Viewer}} 
to parse, display, and plot the data from the FITS files.

Peer reviewed XSM data will be made available to the community after a lock-in period of 
a maximum of nine months from the end of each observing season from the ISRO Science Data Archive
(ISDA)\footnote{\url{https://www.issdc.gov.in/isda.html}} hosted at ISSDC.
While releasing the data, the XSMDAS and CALDB, and associated documentation
will also be provided so that the users can process the data to generate products
according to their specific scientific objectives. Updates to XSMDAS or CALDB
will be released as and when any changes are made.

All the XSM data files can be read by using the FITS libraries available with
most of the programming languages like IDL and python for visualization or 
to carry out further analysis. 
Spectrum and response files of XSM are compatible with the X-ray astronomy
spectral fitting software tools XSPEC and ISIS. The spectra can be directly loaded into 
them following the standard procedures to carry out spectral fitting 
with physical or empirical models.

OSPEX\footnote{\url{https://hesperia.gsfc.nasa.gov/ssw/packages/spex/doc/ospex_explanation.htm}} 
is another software that is widely used
in the solar physics community for X-ray spectral fitting. It is an IDL-based package
available as part of Solarsoft. The time-series calibrated spectra with one-second cadence 
included in the XSM data archive can be loaded into OSPEX with an
IDL routine {\verb ch2xsm_read_data.pro } provided along with the XSMDAS. This routine
loads the XSM spectra and response into the OSPEX data structures. Further manipulation
of the spectra, such as binning to specific time duration, can then be carried out 
with the features present in OSPEX.

\section{Summary}
\label{sec:summary}

A suite of software modules developed for the processing of the data
from XSM instrument on-board the Chandrayaan-2 mission
are presented. These include a user-level analysis software, XSMDAS, 
with functionalities to generate the science products from the XSM data, viz.
light curves and spectra taking into account all aspects of calibration
captured in the calibration database. This software
offers flexibility to the user for selection of all
feasible binning parameters, such as time and energy ranges that suites
their specific requirement and the product files are directly readable
by widely used spectral and temporal analysis tools. The XSMDAS
will be made available along with the XSM raw data and standard calibrated
products.

Apart from this user-level software, a GUI tool named XSMQLD
has also been developed to have a first look at the data and house-keeping information,
which uses the XSMDAS modules as the back-end. XSM has been operating and
acquiring data regularly from September 2019. The data processing at
the Payload Operations Center is completely automated with the
XSMDMS. It also maintains a database to allow easy access and management and an internal
web application for monitoring the data and instrument health. The XSMDMS
is operational at the POC and will continue to handle uninterrupted
data processing for the entire mission duration.

\section*{Acknowledgments}

XSM payload was designed and developed by Physical Research Laboratory (PRL),
Ahmedabad, supported by the Department of Space, Govt. of India with contributions
from U. R. Rao Satellite Centre (URSC), Bengaluru, Laboratory for Electro-Optics
Systems (LEOS), Bengaluru, and Space Application Centre (SAC), Ahmedabad.
We thank various facilities and technical teams of all the above centers
for their support during the design, fabrication, and testing of this instrument.
The Chandrayaan-2 mission is funded and managed by the Indian Space Research Organisation (ISRO).
The authors thank Chandrayaan-2 mission, operations, PACQ and level-0 data processing
teams, and ISSDC team for their support for the instrument operations, data download,
and ground pre-processing.
We also thank the external test and evaluation committee who
carried out the testing of the XSM software and provided suggestions.
Members of the Chandrayaan-2 PDS4 working group are acknowledged for
discussions related to the generation of PDS4 compliant archive.
NPSM thanks Varun Bhalerao (IIT Bombay) and Ajay Vibhute (IUCAA, Pune) 
for valuable discussions on the development of POC data management tools.
The software presented in this work makes use of several open-source libraries, 
and we acknowledge their contributors.

\bibliographystyle{model2-names}
\biboptions{authoryear}
\bibliography{references}

\begin{thebibliography}{6}
\expandafter\ifx\csname natexlab\endcsname\relax\def\natexlab#1{#1}\fi
\providecommand{\url}[1]{\texttt{#1}}
\providecommand{\href}[2]{#2}
\providecommand{\path}[1]{#1}
\providecommand{\DOIprefix}{doi:}
\providecommand{\ArXivprefix}{arXiv:}
\providecommand{\URLprefix}{URL: }
\providecommand{\Pubmedprefix}{pmid:}
\providecommand{\doi}[1]{\href{http://dx.doi.org/#1}{\path{#1}}}
\providecommand{\Pubmed}[1]{\href{pmid:#1}{\path{#1}}}
\providecommand{\bibinfo}[2]{#2}
\ifx\xfnm\relax \def\xfnm[#1]{\unskip,\space#1}\fi
\bibitem[{{Acton}(1996)}]{acton96}
\bibinfo{author}{{Acton}, C.H.}, \bibinfo{year}{1996}.
\newblock \bibinfo{title}{{Ancillary data services of NASA's Navigation and
  Ancillary Information Facility}}.
\newblock \bibinfo{journal}{\planss} \bibinfo{volume}{44},
  \bibinfo{pages}{65--70}.
\newblock \DOIprefix\doi{10.1016/0032-0633(95)00107-7}.
\bibitem[{{Arnaud}(1996)}]{arnaud96}
\bibinfo{author}{{Arnaud}, K.A.}, \bibinfo{year}{1996}.
\newblock \bibinfo{title}{{XSPEC: The First Ten Years}}, in:
  \bibinfo{editor}{{Jacoby}, G.H.}, \bibinfo{editor}{{Barnes}, J.} (Eds.),
  \bibinfo{booktitle}{Astronomical Data Analysis Software and Systems V},
  p.~\bibinfo{pages}{17}.
\bibitem[{{Houck} and {Denicola}(2000)}]{2000ASPC..216..591H}
\bibinfo{author}{{Houck}, J.C.}, \bibinfo{author}{{Denicola}, L.A.},
  \bibinfo{year}{2000}.
\newblock \bibinfo{title}{{ISIS: An Interactive Spectral Interpretation System
  for High Resolution X-Ray Spectroscopy}}. volume \bibinfo{volume}{216} of
  \textit{\bibinfo{series}{Astronomical Society of the Pacific Conference
  Series}}.
\newblock p. \bibinfo{pages}{591}.
\bibitem[{{Mithun} et~al.(2020){Mithun}, {Vadawale}, {Shanmugam}, {Patel},
  {Tiwari}, {Adalja}, {Goyal}, {Ladiya}, {Singh}, {Kumar}, {Tiwari}, {Modi},
  {Mondal}, {Sarkar}, {Joshi}, {Janardhan} and {Bhardwaj}}]{mithun20}
\bibinfo{author}{{Mithun}, N.P.S.}, \bibinfo{author}{{Vadawale}, S.V.},
  \bibinfo{author}{{Shanmugam}, M.}, \bibinfo{author}{{Patel}, A.R.},
  \bibinfo{author}{{Tiwari}, N.K.}, \bibinfo{author}{{Adalja}, H.L.},
  \bibinfo{author}{{Goyal}, S.K.}, \bibinfo{author}{{Ladiya}, T.},
  \bibinfo{author}{{Singh}, N.}, \bibinfo{author}{{Kumar}, S.},
  \bibinfo{author}{{Tiwari}, M.K.}, \bibinfo{author}{{Modi}, M.H.},
  \bibinfo{author}{{Mondal}, B.}, \bibinfo{author}{{Sarkar}, A.},
  \bibinfo{author}{{Joshi}, B.}, \bibinfo{author}{{Janardhan}, P.},
  \bibinfo{author}{{Bhardwaj}, A.}, \bibinfo{year}{2020}.
\newblock \bibinfo{title}{{Ground Calibration of Solar X-ray Monitor On-board
  Chandrayaan-2 Orbiter}}.
\newblock \bibinfo{journal}{Experimental Astronomy (submitted)} ,
  \bibinfo{pages}{arXiv:2007.07326}\href{http://arxiv.org/abs/2007.07326}{\tt
  arXiv:2007.07326}.
\bibitem[{{Radhakrishna} et~al.(2020){Radhakrishna}, {Tyagi}, {Narendranath},
  {Vadodariya}, {Yadav}, {Singh}, {Balaji}, {Satya}, {Shetty}, {Suresh Kumar},
  {Kumar}, {Vaishali}, {Pillai}, {Tadepalli}, {Raghavendra}, {Sreekumar},
  {Agarwal} and {Valarmathi}}]{radhakrishna20}
\bibinfo{author}{{Radhakrishna}, V.}, \bibinfo{author}{{Tyagi}, A.},
  \bibinfo{author}{{Narendranath}, S.}, \bibinfo{author}{{Vadodariya}, K.},
  \bibinfo{author}{{Yadav}, R.}, \bibinfo{author}{{Singh}, B.},
  \bibinfo{author}{{Balaji}, G.}, \bibinfo{author}{{Satya}, N.},
  \bibinfo{author}{{Shetty}, A.}, \bibinfo{author}{{Suresh Kumar}, H.N.},
  \bibinfo{author}{{Kumar}}, \bibinfo{author}{{Vaishali}, S.},
  \bibinfo{author}{{Pillai}, N.S.}, \bibinfo{author}{{Tadepalli}, S.},
  \bibinfo{author}{{Raghavendra}, V.}, \bibinfo{author}{{Sreekumar}, P.},
  \bibinfo{author}{{Agarwal}, A.}, \bibinfo{author}{{Valarmathi}, N.},
  \bibinfo{year}{2020}.
\newblock \bibinfo{title}{{Chandrayaan-2 Large Area Soft X-ray Spectrometer}}.
\newblock \bibinfo{journal}{Current Science} \bibinfo{volume}{118},
  \bibinfo{pages}{219--225}.
\newblock \DOIprefix\doi{10.18520/cs/v118/i2/219-225}.
\bibitem[{{Shanmugam} et~al.(2020){Shanmugam}, {Vadawale}, {Patel}, {Adalaja},
  {Mithun}, {Ladiya}, {Goyal}, {Tiwari}, {Singh}, {Kumar}, {Painkra},
  {Acharya}, {Bhardwaj}, {Hait}, {Patinge}, {Kapoor}, {Kumar}, {Satya},
  {Saxena} and {Arvind}}]{shanmugam20}
\bibinfo{author}{{Shanmugam}, M.}, \bibinfo{author}{{Vadawale}, S.V.},
  \bibinfo{author}{{Patel}, A.R.}, \bibinfo{author}{{Adalaja}, H.K.},
  \bibinfo{author}{{Mithun}, N.P.S.}, \bibinfo{author}{{Ladiya}, T.},
  \bibinfo{author}{{Goyal}, S.K.}, \bibinfo{author}{{Tiwari}, N.K.},
  \bibinfo{author}{{Singh}, N.}, \bibinfo{author}{{Kumar}, S.},
  \bibinfo{author}{{Painkra}, D.K.}, \bibinfo{author}{{Acharya}, Y.B.},
  \bibinfo{author}{{Bhardwaj}, A.}, \bibinfo{author}{{Hait}, A.K.},
  \bibinfo{author}{{Patinge}, A.}, \bibinfo{author}{{Kapoor}, A.h.},
  \bibinfo{author}{{Kumar}, H.N.S.}, \bibinfo{author}{{Satya}, N.},
  \bibinfo{author}{{Saxena}, G.}, \bibinfo{author}{{Arvind}, K.},
  \bibinfo{year}{2020}.
\newblock \bibinfo{title}{{Solar X-ray Monitor (XSM) On-board Chandrayaan-2
  Orbiter}}.
\newblock \bibinfo{journal}{Current Science} \bibinfo{volume}{118},
  \bibinfo{pages}{45--52}.
\newblock \DOIprefix\doi{10.18520/cs/v118/i1/45-52},
  \href{http://arxiv.org/abs/1910.09231}{\tt arXiv:1910.09231}.

\end{thebibliography}

\end{document}